# Preprint A Game Based Assistive Tool for Rehabilitation of Dysphonic Patients

Zhihan Lv*
Fundacion Instituto Valenciano de Neurorrehabilitacion (FIVAN), Valencia, Spain
Chantal Esteve†
Fundacion Instituto Valenciano de Neurorrehabilitacion (FIVAN), Valencia, Spain
Javier Chirivella
Fundacion Instituto Valenciano de Neurorrehabilitacion (FIVAN), Valencia, Spain
Pablo Gagliardo‡
Fundacion Instituto Valenciano de Neurorrehabilitacion (FIVAN), Valencia, Spain

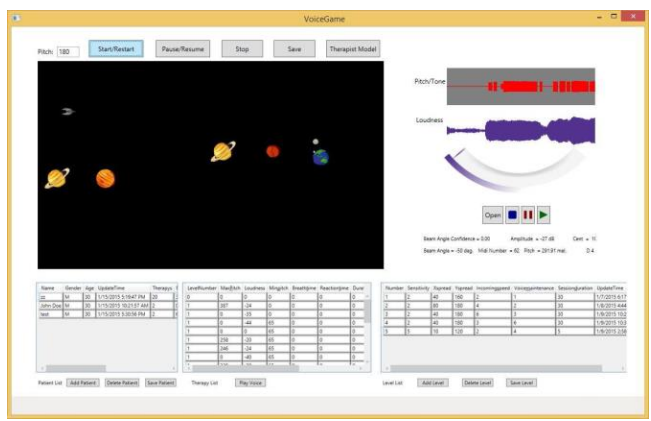

Figure 1: The full model of the system running scene.

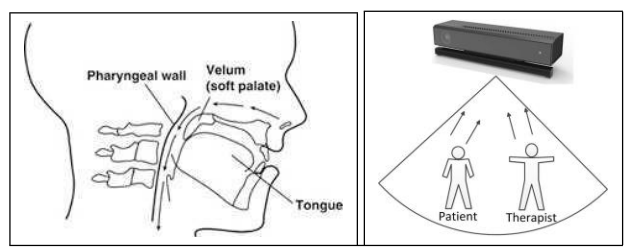

Figure 2: Left: Vocal muscles; Right: The system schematic diagram.

## ABSTRACT

An assistive training tool for rehabilitation of dysphonic patients is designed and developed according to the practical clinical needs. The assistive tool employs a space flight game as the attractive logic part, and microphone arrays as input device, which is getting rid of ambient noise by setting a specific orientation. The therapist can guide the patient to play the game as well as the voice training simultaneously side by side, while not interfere the patient voice. The voice information can be recorded and extracted for evaluating the long-time rehabilitation progress. This paper outlines a design science approach for the development of an initial useful software prototype of such a tool, considering 'Intuitive', 'Entertainment', 'Incentive' as main design factors.

**Index Terms:** 1.3.7 [Computer Graphics]: Three-Dimensional Graphics and Realism—Virtual reality

## 1 INTRODUCTION

Virtual reality and augmented reality have been applied in a wide range of topics in healthcare [50]. Using video game as the vision feedback has been already demonstrated as a useful HCI approach for rehabilitation assistive method [19], both in motor [9] and cognitive [17] [3]. In the technology trend, vocal interaction research has slowly been gaining popularity in the mainstream HCI, assistive technology, arts and game development communities [16]. Especially, using voice as input interface is considered as a powerful system control technique in some inspiring research [37] [18]. Although the current most intuitive input method is still motion sensing interaction [26], such as hands interaction [28], feet interaction [21], head interaction [14] and multimodal interaction [27] embeded in the wearable devices [23] like smart glasses [25]. However, the people with disabilities may not be able to freely control their hands or feet during the rehabilitation exercises, even not be able to move independently leaving wheelchair [8]. Previous research have proved the usability of the voice interactive serious game in speech therapy [40], for children with autism [13], and parkinsons disease patients [15].

Reducing costs continuously become the pressure of contemporary healthcare systems. The total number of patients will increase while the available healthcare staff will decline in the coming decade [32] [39]. Technological advancement makes computer software to be viable and affordable alternatives to provide assistive functions for long-time rehabilitation and to reduce the workload of care professionals. Nevertheless, in order to make sure that implement assistive tool will not dehumanize healthcare, friendly user interactive experience is a must for this technology; what's more, according to the specific circumstances of patient, this technology should be able to be flexible customized by therapists. During interaction, these user interactive capabilities allow the game based assistive tool actively engage the patient and arouse enthusiasm. Due to patients' daily verbal communication, compared to non-interactive technology, the assistive tool has potential to be better at motivating users and inducing voice improvement. In addition, by embedding voice pitch estimation methodology, users can use their current voice rating to interact with the game. For patients that have difficulty in making healthful-human voice, using current voice pitch becomes important. However, because most patients never encountered a rehabilitation process of using serious game based assistive tool before, they know little about what kind of serious games could enrich their therapy and how to design them so that they are accepted in long-time rehabilitation process.

---

*e-mail: lvzhihan@gmail.com, zlu@neuroathome.com
†e-mail: chantal@fivan.org
‡e-mail: pablog@fivan.org

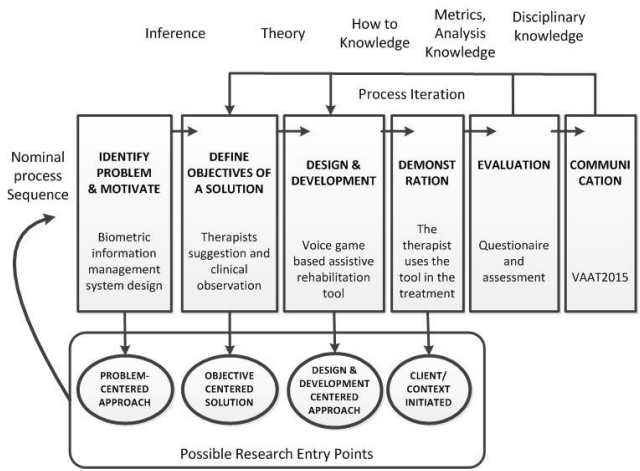

Figure 3: Adapted DSPM for our research.

In our research, we employ video game as vision feedback tool to evaluate the efficacy of the patient voice exercise. The efficacy of the patient voice exercise can be considered as the long-time continued evaluation of the severity of dysphonia. Our application software detects the pitch and loudness as the evaluation factor. Thank to the microphone arrays of kinect which is enabled to estimate the location of the sound source, our software support filtering ambient noise, so that the therapist can supervise and train the patient side by side. In the game logic design, 'Intuitive', 'Entertainment', 'Incentive' are three main factors considered in our design principle.

We define our research just like a biometric information management system design, in which the biometric refers to voice pitch specifically, while serious games are employed as the user interface. Solving a previously unsolved problem [11] by using a design science (DS) approach is the significant way of the novel research [1] [46]. Recently, while DS has mostly been discussed in information systems research [11] [4] [33] [31], it is equally applicable for human-computer-interaction research projects [42] [2] [7] and computer game [5] [10]. In our research, we chose the design science process model (DSPM) [33] and the process adapted to the research project was shown in Figure 3. This paper's reminder is structured according to the DSPM. Because the design of the tool was motivated by an identified problem, the research can be classified as a problem centered initiation. The activities problem identification and motivation, definition of objectives of a solution, and design and development were conducted in sequential order. Figure 3 demonstrate these steps in the whole design loop process. In this paper, we just start the activities demonstration and the evaluation have not been carried out completely yet. In the preliminary evaluation, suggestions inspired from how we have gone forward during those activities are discussed. The publication of this paper is considered as the beginning of the communication activity.

## 2 THERAPISTS SUGGESTION AND CLINICAL OBSERVATION

In clinical, dysphonia is measured using a variety of tools that allow the clinician to see the pattern of vibration of the vocal folds, principally laryngeal videostroboscopy, such as coustic examination and electroglottography. Subjective measurement of the severity of dysphonia is carried out by trained clinical staff (e.g. GRBAS (Grade, Roughness, Breathiness, Asthenia, Strain) scale, Oates Perceptual Profile). Objective measurement of the severity of dysphonia typically requires signal processing algorithms applied to acoustic or electroglottographic recordings [20]. Besides invasive or drug based treatments, effective logopedic treatments have been proved [36]. However, constant training is a key factor for this type of therapy [15].

In theory, the tone of pronunciation depends on the mouth and facial muscles coordination. Therefore, the amount of dysphonic patients is much smaller than others limbs disable patients. In our sample database, it's one fourth (fifteen of sixty), in which four or five are expected to be recovered with computer games as assistive tools. According to our sample database, the dysphonic voice can be hoarse or excessively breathy, harsh or rough, but some kind of phonation is still possible, therefore, the clear speech recognition is not expected. The patients usually have inabilities of pronunciation of the high pitches. The hopeful assistive tool is able to retrieve the quantified long-time voice rehabilitation information. Dysarthria and dysphagia occur frequently in Parkinsons disease [44]. The voice exercise can not only improve the pronunciation but also enhance the swallowing muscle (Velum and Pharynx, see figure 2 left) of the patient, in case the patient eat the food into lung accidentally.

In order to identify and define the problem pragmatically, we discussed with a therapist regarding general exercise process. The therapist is the admitted domain expert, who has rich experience in voice rehabilitation training, in addition, she is a semi-professional singer and knows deep understand about the principle of human voice. The most important is, she has a positive attitude for the possibility of voice rehabilitation using game as assistive tool. She brings us some inspiring suggestions and introduces the general exercises what are usually used in the clinical rehabilitation treatment. After the discussion, she presents three reality treatments to show us the potential scenarios to apply our designed work.

### 2.1 General Exercise

**Exercise (1)** Breathing is an important issue during the exercise, the therapists have to teach the patients to control breath before exercise to avoid the patients to choke.

**Exercise (2)** Using two hands to pull each other or up-legs can stretch the muscle so that enhancing the tone and loudness.

**Exercise (3)** The therapists speak 'CRAC' in various of tones and then ask the patient to follow. The patient need to reach the specific tone in required seconds, and also maintain in a specific tone for required time.

**Exercise (4)** Both singing and speaking are good methods for rehabilitation exercise. So singing is an available choice as long as the patient is able to vocalize coherent voice.

**Exercise (5)** Number (i.e. one, two, three, ) and date (i.e. Monday, Tuesday, ) are also reasonable material since they are the most common words in the life. The reading exercise is iterative. The patients read only one word at the beginning, and then increase the number of the words gradually, till they can talk a whole sentence.

**Exercise (6)** The therapists used to record the pronunciation of patients so that comparing the difference and indicating the progress. Both the real-time value and maximum value of the tone of the patient voice need to be presented and recorded.

### 2.2 Clinical Observation

Followed with the therapist, we have met three volunteer dysphonic patients and observed their treatments, according to the agreement with Declaration of Helsinki [45].

**First Case**

The first patient was physical weakness and sitting in the wheelchair, it took nearly 5 minutes. He cannot conduct exercise (2) while he struggle to speak using his facial muscle. The therapist guided exercise (3) and (5). The patient followed it until the tone was too high to pronounce for him.

**Second Case**

The second patient conducted four exercises, it took nearly twenty minutes.

1. Lying on the bed, did exercise (1) for twenty seconds. At the same time the therapist was counting. This exercise was not easy for the patient because it made her sleepy.
2. Sat and did exercise (2) and (3) and the same time.
3. Stood against the wall, squatted and stood slowly, and did exercise (5).
4. Sat and read about 50 listed words with [rr] pronunciation. Afterward, read short sentences with five words, and then long sentences.

**Third Case**

The third patient is a Parkinson's patient. This training takes 5 minutes. The patient sat on the bed and did exercise (5). The patient voice was obviously mono. The pitch of the therapist and the patient were measured by a professional tool. The pitch value of the therapist was static on 60, while that of the patient was range from 51 to 56, which cannot reach the required level.

## 3 SYSTEM IMPLEMENTATION

### 3.1 System User Interface

According to the therapists suggestions and our clinical observation, considering the design factors simultaneously, we designed an serious game based assistive tool with a user interface, the full mode (Therapist Mode) is as in figure 1. In this mode, the game logic (top-left), voice analysis (top-right), patient case history (bottom-left) and the game level editor (bottom-right) are embedded in the main window. The patient case history and the level editor are implemented by database management user control, it's rapid-developed and modifiable which is suitable for our gradual research and iterative development. The voice analysis UI visualizes pitch (red), loudness (purple) by dense histogram, and voice source by a semicircular compass. Through setting the voice source direction on the semicircular compass, only the voice spreaded from the patient location could be received by the system, while all of noised or other interference voice including therapist voice will be filtered.

### 3.2 Biometric Estimation

The software is able to estimate the basic information of patient voice (i.e. pitch (Mel) and loudness (dB)). Further more, the midi number and note name are calculated. The frequency is identified by FFT (Fast Fourier Transform) based frequency-domain approach which turns the voice wave into a frequency distribution. Further more, the pitch (tone) and midi number are calculated depend on frequency, as in formula:

$$Pitch = 1000 \times \log_2(1 + f); \quad (1)$$

$$Midi = 69 + 12 \times \log_2(f/440); \quad (2)$$

Where 440 is the absolute pitch frequency. The real-time frequency value is visible on the UI in the light of therapist needs. Generally, the normal range of pitch is 80 Mel to 200 Mel for man and 150 Mel to 350 Mel for woman. The pitch extracting method was evaluated by the therapist with a piano software, which is a reliable tool for therapists to present the accurate tones (Alias: pitch). The evaluation result indicates the accurate rate is enough high for our demand. Considering the voice characteristics like non-continuity and low-loudness of dysphonic patients, the sensitivity and suitability of the pitch extracting method are the decisive factors beside the accuracy rate. Therefore we employed FFT based pitch estimation algorithm instead of auto-correlation based time-domain approach [34] which is not sensitive or accurate enough for the voice of dysphonic patients.

### 3.3 Game Interaction

Our rich clinical experience gives rise to unique game design principle. The entertainment is significant for the designed game, because it can stimulate the initiative and enhance the motivation of the patient so that the patients are willing to pay more effort to the rehabilitation exercise. 2D game is enough attractive, because the patients we have met often confuse the 3D manipulation rule. Besides, some of the patients have disability of 3D perception. Therefore, aiming to our practical demand, 2D UI is prefered and easier to navigate, traverse and engage, which satisfies criteria that arise from the nature of patient's spatial perception [47].

As suggested by the therapist, pitch (Mel) is the main controlable factor, thus we uses it as the manipulating tool which function is equal to the joystick in the classic video game. Meanwhile, the game UI renders the vision feedback of the user manipulation. Some other factors are also considered, such as Phonation time (ms), pitch change (Mel), Duration (s), Reaction time (ms).

**Phonation Time:** The maximum value of the phonation time.

**Pitch Change:** The maximum value and the minimum non-zero value of patient's pitch.

**Duration:** How long the patient hold on in the game.

**Reaction Time:** This value is measured by patient's reaction time of the incoming obstacle.

Built based on the considered factors, the configuration includes some essential parameters which can modify the difficult level of the games. The parameters contain sensitivity, x spread, y spread, incoming speed, voice maintenance, session duration. The parameters are available to modify by the level editor, and the new level will be generated for the specific patients as soon as it's saved.

**X Spread, Y Spread:** The initial position of the spaceship.

**Incoming Speed:** The moving speed of the flying obstacles.

**Voice Maintenance:** It depends the width of the obstacles. The patients need to keep high-enough pitch so that the spaceship can maintain the altitude and leap the obstacles.

**Session Duration:** The duration of the level. If game over in the duration time, the player fail.

The designed functions can provide immediate evaluation feedback to therapists and patients. One space flight game is created as the logic part of the assistive tool, as shown in figure 1. In this game, the player uses voice pitch to control the spaceship as the avatar to dodge the planets. The spaceship moves up if the pitch is higher than 200, and move down if lower than it, the value is adjustable. Once the spaceship collides the planets, the game is over and the player score is recorded. Considering the manipulation complexity for the dysphonic patients, the loudness factor isn't utilized in the game interaction.

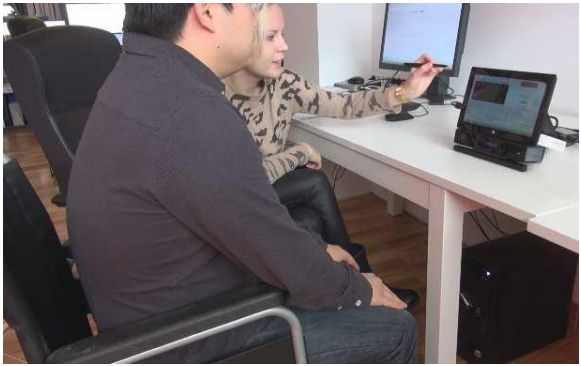

Figure 4: Running scene.

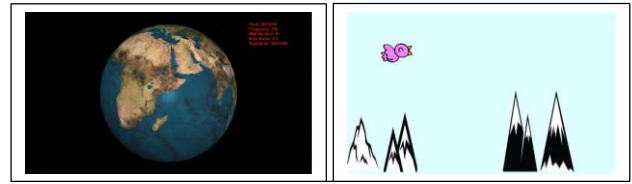

Figure 5: Left: The EARTHSCREAM game prototype; Right: The BIRDFLY game prototype.

### 3.4 Assisting Functions

Beside the designed game stories, some assisting functions are implemented, include 'create patient', 'record and replay voice', 'save therapy'. In our gamification therapy, one game session is one therapy, that means when the therapist starts a game session for patient, the therapy starts synchronously.

**Create Patient** Create the case history for patients. Beside the patients personal information, both game results and therapy information are saved in the case history.

**Record and Replay Voice** Record and replay the voice of the patients during the treatment, so that the therapists can analysis them as the valuable material for a long-time rehabilitation.

**Save therapy** Save all the estimated information in the treatment, such as maximum pitch, maximum loudness, breath time, reaction time.

### 3.5 Running Scene

When the software is running, both therapist and patient stand in front of Kinect2 and watch the monitor as in figure 4. The kinect2 can recognize the patient by location. During the game, the patient will control the game avatar to follow the game role, at the same time, the therapist makes voice to supervise the patient. The implied gamification feature during this process is that the patient try the best to follow the therapist's voice guide. When the game is over, the software will save the biometric data automatically.

The system currently runs on the Microsoft Surface with Windows 8 smoothly. The software is developed based on Windows Presentation Foundation (WPF) library with C# programming language. Thanks to the portability of WPF, the current system has potential to be ported to both mobile device and 4K-TVs with popular portable operating system, such as Android and IOS. We also plan to employ portable head mounted device (HMD) (e.g. google cardboard, vrAse) to enhance the immersion of the game for patients focusing on the therapy. The portable version is also suitable for rehabilitation@home application [6].

## 4 PRELIMINARY EVALUATION

This paper outlined the motivation for the development of a game based assistive tool for rehabilitation of dysphonic patients, defined objectives for its construction and sketched a high level view of the design and technology choices. While this is a good start point of developing practical software for clinical application and evaluating the usability preliminarily. The construction of the first version of the proposed assistive tool has already revealed some defects and ideas for further improvements. Thus the next step is to iteratively improve it and to develop experiments to test it before eventually moving on to conducting clinical tests as the assistive tool for rehabilitation of dysphonic patients. The suggestions from domain experts in clinical treatment are needed to achieve this.

Since most test subjects (i.e. patients) will not have experienced these games as the assistive tool, it is hard to measure how accurate the voice control input actually was as well as how funny the serious game actually was. The most obvious approach is to involve domain experts and rely on their feedback, here the domain experts are the therapists. Another alternative is developing a generic questionnaire to evaluate the usability of the assistive tool.

As an extreme counter-example, we have developed a 3D game prototype 'EARTHSCREAM' using XNA game engine as shown in figure 5 left, in order to verify the usability of 3D game and further compare it with 2D game. A rotating 3D earth presents both pitch and loudness information. The speed of rotation represents the pitch. The pitch intermediate value is 200 which decides the earth rotation orientation, toward left if less than 200, otherwise right. The rotation speed depends on the absolute value of the difference between the current pitch value and the pitch intermediate value. Here we use the cloud layer strength to represent the loudness. The higher loudness, the darker cloud layer is rendered. However, the therapist declines to use it for patients, she explains that as most of her patients are disable to understand this unintuitive game rule although it is uncomplicated. Therefore, 'Intuitive' is one of the most important factor beside entertainment for rehabilitation game design.

It's also worth to mention that the therapists demands need to be developed along with the software development. When we discussed with the therapist regarding the clinical demand of this assistive tool at the beginning, the therapist proposed a very simple game design. The game rule is using voice pitch to control a height of a ball to hit a oncoming ball. Obviously it conforms the 'Intuitive' design principle, however, it definitely lacks entertainment. Therefore, we proposed several more entertaining suggestions as we described in this paper, which could substitute the therapist's original simple game logic. Delightedly the therapist accepted our suggestions, even brought more professional ideas to supplement. In this way, our research is conducted gradually, at the same time the assistive tool software is constructed iteratively till the therapist show her satisfaction.

During the gradual and iterative process, another design principle attracts our attention inadvertently, that is 'Incentive'. 'Incentive' is usually the source of entertainment, since it stimulates the player to struggle to earn the most award. However, cheating cannot be ignored in all kinds of games absolutely, although in the serious games. Otherwise, the meticulous-designed game rule is damaged, meanwhile, the original motivation of rehabilitation utility is missing. In the space flight game, the planets are generated at random start points for the purpose of fulfilling the unpredictability design principle. Nonetheless, the randomly generated planets cannot spread all the start points, thus one or more 'always safe' position exist inevitably, which causes the cheating of patients, to win the game as long as staying at those positions without any ef-fort. This kind of cheating is avoided by attachment of a permanent gravity on the avatar. We developed a BIRDFLY game to express this design idea, as shown in figure 5 right. In this game, the patients are forced to make voice to counterwork the gravity and keep the bird avatar flying in sufficient height, thus avoidance of drop to the obstacle at bottom.

## 5 CONCLUSION

Virtual reality has been widely applied in many fields range from everyday entertainment and communication [24] [48] to traditional research fields such as geography [29] [38] [30] [41] and biology [43] [12]. In this paper, we introduced a virtual reality technology based application in clinical rehabilitation field.

We developed a voice interactive serious game as a likely solution for providing assistive rehabilitation tool for therapists. The recorder of the voice of patients could be analysed to evaluate the long-time rehabilitation results, as well as the prediction of the rehabilitation process. In addition, the controllability and the happiness of the patients are the two most important factors of the usability of the designed rehabilitation games. Therefore, we plan to evaluate the emotion, which includes the happy extent as well as the control extent. Emotion is a significant part of users decision-making ability, which can evaluate the product against different emotions that a user goes through. In order to assess the patients' emotions, we plan to use the Geneva Emotions Wheel (GEW) [35]. GEW allows us to address the pleasantness and control dimensions of emotions. Rhythm and tempo controlling is an interesting ability which can be measured by the metronom. We also consider to concentrate patient' attention by enhancing the vision feedback using stereoscopic 3D video [49] or pseudo 3D [22] technolgoy.

## ACKNOWLEDGMENT

The authors would like to thank Sonia Blasco and Vicente Penades for their fruitful help and suggestions, and Oktawia Karkoszka for recording the demo video. The work is supported by LanPercept, a Marie Curie Initial Training Network funded through the 7th EU Framework Programme under grant agreement no 316748.